# Possible observation of parametrically amplified coherent phasons in $K_{0.3}MoO_3$ using time-resolved extreme-ultraviolet ARPES

H. Y. Liu[1,*], I. Gierz[1], J.C. Petersen[1,2], S. Kaiser[1], A. Simoncig[1], A.L. Cavalieri[1], C. Cacho[3], I.C.E. Turcu[3], E. Springate[3], F. Frassetto[4], L. Poletto[4], S.S. Dhesi[5], Z.-A. Xu[6], T. Cuk[7], R. Merlin[8] and A. Cavalleri[1,2,‡]

[1]*Max Planck Department for Structural Dynamics, CFEL, Hamburg, Germany*
[2]*Department of Physics, Clarendon Laboratory, University of Oxford, UK*
[3]*Central Laser Facility, STFC Rutherford Appleton Laboratory, UK*
[4]*CNR-Institute for Photonics and Nanotechnologies, Padova, Italy*
[5]*Diamond Light Source Ltd., Chilton, UK*
[6]*Department of Physics, Zhejiang University, Hangzhou, China*
[7]*Materials Sciences Div. and Physical Biosciences Div., Lawrence Berkeley National Laboratory, University of California, Berkeley*
[8]*Department of Physics, University of Michigan*

*haiyun.liu@mpsd.cfel.de
‡andrea.cavalleri@mpsd.cfel.de

**Abstract:** We use time- and angle-resolved photoemission spectroscopy (tr-ARPES) in the Extreme Ultraviolet (EUV) to measure the time- and momentum-dependent electronic structure of photo-excited $K_{0.3}MoO_3$. Prompt depletion of the Charge Density Wave (CDW) condensate launches coherent oscillations of the amplitude mode, observed as a 1.7-THz-frequency modulation of the bonding band position. In contrast, the anti-bonding band oscillates at about half this frequency. We attribute these oscillations to coherent excitation of phasons via parametric amplification of phase fluctuations.

The total energy of a low-dimensional metal can be lowered by a periodic distortion of the crystal lattice. Such a Peierls transition enlarges the unit cell and redistributes charge density, opening band gaps at the Fermi level [1]. The resulting Charge Density Wave (CDW) ground state exhibits new low-energy collective excitations, the *amplitudon* and *phason*, which correspond to distortions and translations of the modulated charge density [2].

The amplitude mode is weakly momentum-dependent with a finite frequency at $q = 0$. In contrast, the phase mode energy increases with momentum q, dispersing linearly near the Brillouin zone center. In the case of non-commensurate CDWs, the phase mode has ideally zero energy at $q = 0$, corresponding to zero DC resistance. However, pinning to defects produces a gap in the phase-mode spectrum of most materials and the phase mode frequency is non-zero at $q = 0$ [3].

A widely studied quasi-one-dimensional (1D) CDW material is the linear chain compound $K_{0.3}MoO_3$, known as blue bronze [4, 5]. Below $T_{CDW}$ = 180 K a CDW forms, and a gap opens at the Fermi level [6, 7]. New collective excitations appear in the infrared, Raman, and neutron spectra [8, 9, 10, 11]. The amplitude mode lies at 1.7 THz, and softens with increasing temperature [11, 12]. The $q = 0$ phase mode is pinned and its frequency is 0.1–0.2 THz [11, 13].

Femtosecond optical pulses can be used to trigger rearrangements in the collective properties of this and other complex solids [14]. The photo-induced dynamics driven by ultrashort optical pulses result from strong perturbations of the ground state, and proceed along physical pathways that are not easily predicted by the linear response theory used to describe fluctuations of the ground state. For example, intense photo-excitation can readily destroy charge gaps [15, 16], and the underlying coherent pathways can only be partially understood by considering the near equilibrium normal modes of the solid [17, 18].

Momentum- and time-dependent electronic structural dynamics can be measured by time- and angle-resolved photoemission spectroscopy (tr-ARPES). This method has already been applied to quasi-two-dimensional (2D) CDW compounds such as 1$T$–TaS$_2$ [19, 20], TbTe$_3$ [21] and 1$T$–TiSe$_2$ [22]. CDW gaps of various types were seen to melt [23], and Raman-active amplitude modes were excited [24].

In our experiment, 30 fs pulses at 790-nm wavelength were used to stimulate the sample with a fluence of 0.5 mJ/cm$^2$. Synchronized EUV pulses with a photon energy of 20.4 eV were generated by High-order Harmonic Generation (HHG), filtered by a time preserving monochromator and used to probe the sample by angle-resolved photoemission spectroscopy (ARPES) [20, 25]. The kinetic energies and emission angles of the photoelectrons provide snapshots of the electronic structure. The energy resolution of our experiments was about 150 meV, mainly limited by the EUV source characteristics. The combination of high photon energy and spectral resolution makes it possible to study a series of small gaps throughout the large Brillouin zone of blue bronze. By varying the delay between the excitation and the EUV pulses, we record the electronic structure as it evolves in time.

The crystal structure of blue bronze features 1D chains of MoO$_6$ octahedra (Fig. 1(a)), which extend along the $b$ direction, arranged side-by-side in slabs that are separated from one another by sheets of K ions [26]. The material can be cleaved in the plane of the slabs, and the chains then lie along the surface plane. In our experiment, we measured photoelectrons with momentum along the chains, which corresponds to the Γ-Y-X' direction in reciprocal space. The pump and probe beams are polarized in that direction as well. The sample, grown as described in Ref. 27, is cleaved and measured under ultra-high vacuum at a temperature of 20 K.

The unperturbed band structure was obtained from photoemission at time delays before the excitation pulse. Figure 1(b) shows four bands reaching the Fermi level at $k_F^{(A1)} = 0.6$, $k_F^{(B1)} = 0.9$, $k_F^{(A2)} = 1.1$, and $k_F^{(B2)} = 1.25$, in units of $\pi/b$, the distance to the zone boundary. These anti-bonding (A1 and A2) and bonding (B1 and B2) bands result from the hybridization of Mo-4$d$ and O-2$p$ states [28, 29]. Their dispersion agrees with calculations [28, 29] and previous ARPES experiments [4, 30, 31, 32]. Figure 1(c) shows individual energy distribution curves (EDCs) for each momentum. The minimum binding energy is 0.15–0.2 eV, due to the opening of the CDW gap.

EDCs at $k_F^{(B1)}$ and $k_F^{(A2)}$ cross a single band, but those at $k_F^{(A1)}$ and $k_F^{(B2)}$ cross both a bonding and an anti-bonding band, giving double-peak features.

In the pump-probe experiments reported here, the optical pulse creates quasiparticles across the gap, depleting the CDW condensate. This is reflected in the changes of the spectral function presented in Fig. 2(a). In each band, an overall upward shift of intensity is observed as the CDW gaps close. Figure 2(b) illustrates the reduction of the CDW gaps by showing individual EDCs at $k_F^{(B1)}$ and $k_F^{(A2)}$, with intensity shifting from the equilibrium valence bands to $E_F$ as the CDW ground state melts non-thermally [33].

By measuring the time-dependent EDCs near $E_F$ for each Fermi momentum, two-dimensional data sets in energy and time are obtained. These are depicted by the density plots in Fig. 3. The prompt initial upward shift, reflecting a reduction of the equilibrium CDW gap, is followed by an oscillation of the band edge [34]. The traces (red and blue data points in Fig. 3) represent the integrated intensity for a 150 meV interval around $E_F$. We observe a fast intensity oscillation in the gap of the B1 band and a slower one in the gap of the A2 band, that we fit to a sine function, undamped within this observation window, with the addition of an exponentially decaying background (black lines in Fig. 3).

The fast mode in Fig. 3(a) has a frequency of 1.7±0.2 THz, consistent with coherent amplitude mode dynamics seen in the past [33, 34]. The slower oscillation of the anti-bonding band (Fig. 3(b)) at 0.8±0.2 THz, observed here for the first time, must indicate the excitation of a different mode. Figures 3(c) and (d) show the dynamics at the other Fermi momenta, $k_F^{(A1)}$ and $k_F^{(B2)}$. There, both the bonding and the anti-bonding bands simultaneously contribute to the signal, and oscillations at both frequencies of 1.7 and 0.8 THz are needed to fit the data.

The observation of 0.8 THz oscillations is surprising, since all Raman-active optical phonons that can be driven by an impulse lie above 2 THz [33]. Only acoustic branches and the phase mode of the CDW are found at lower frequencies [11]. Optical measurements in blue bronze have been interpreted through the excitation of slow phase-mode oscillations near $q = 0$ after [35] suggesting that the phase of the CDW state may in fact be perturbed in that material.

The most likely explanation for the low-frequency oscillation observed in our measurements is the excitation of large-wavevector coherent phase modes, seeded by fluctuations and parametrically amplified by anharmonic coupling to the amplitude mode. This physical mechanism is discussed below and illustrated in Fig. 4.

In Fig. 4(a) the free energy of an incommensurate 1D CDW is shown as a function of the complex order parameter $|\Delta|e^{i\phi}$ [2]. The modulus of the order parameter $|\Delta|$ is, at equilibrium, proportional to the density of the ground state CDW condensate. For a non-commensurate CDW and in absence of pinning the free energy does not depend on the phase $\Phi$ of the order parameter, which is allowed to rotate freely and ideally gives rise zero DC resistance. In blue bronze, quasi-commensurability or pinning to defects fix the phase to a specific value.

In our experiment, optical depletion of the CDW condensate causes an instantaneous reduction of the gap value. Thus, amplitude oscillations are launched about the new equilibrium position, resulting in oscillations of the spectral function in the vicinity of the Fermi level. Photo-excitation also causes de-pinning of the CDW, which in the excited state is free to rotate.

To explain the low frequency *coherent* oscillations we first recall an analogous mechanisms in lattice dynamics, where anharmonicity causes $q = 0$ optical phonons to decay into pairs of acoustic phonons at $\pm q_P$ ($q_P > 0$), conserving both energy and momentum [36], as shown in Fig. 4(b).

For the present case, we consider the anharmonic interaction between amplitude and phase, described by a term of the form $H_A = -\sum_{\alpha,q} \Lambda_{\alpha q} Q_\alpha U_q U_{-q}$, where $\Lambda_{\alpha q}$ are constants, $Q_\alpha$ is the amplitude of the amplitudon at $q = 0$, and $U_q$ is that of the phason of wavevector $q$. If the amplitude mode is coherent, one can recast this anharmonicity as a flow of energy between classical oscillators [37]

$$\ddot{U}_{+q} + \omega_p^2 U_{+q} = \Lambda_{\alpha q} Q_\alpha U_{-q} = \Lambda_{\alpha q} Q_\alpha^{(0)} \sin(\omega_A t) U_{-q}$$
$$\ddot{U}_{-q} + \omega_p^2 U_{-q} = \Lambda_{\alpha q} Q_\alpha U_{+q} = \Lambda_{\alpha q} Q_\alpha^{(0)} \sin(\omega_A t) U_{+q}$$

In these coupled equations $U_{\pm q}$ are the displacements of the phase mode at wavevector $\pm q$. The time-dependence of the impulsively excited amplitudon is described by $Q_\alpha^{(0)} \sin(\omega_A t)$. In Fig. 4(c) we plot a numerical solution of the coupled differential equations given above in the presence of a suitable damping term and phase-mode fluctuations with a white spectrum, which ensures that the initial value for either $U_{+q}$ or $U_{-q}$ is nonzero.

In the color plots of Fig. 4(c), we plot the value of the phason coordinate as a function of time for different ratios of amplitude- and phase-mode frequency. We observe a resonance at $\omega_p = \omega_A/2$, which reinforces the notion that the phase mode is excited at $+/-q_P$. In the 1D lineouts of the phase, we show how at $\omega_p = \omega_A/2$ long lived phase oscillations are observed, whereas away from the resonance condition these oscillations are overdamped.

As a final step to support our line of interpretation, we explain the effect of phase mode oscillations on the band structure. We note that according to textbook CDW physics [2], the wavevector at which nesting causes opening of the CDW gap depends on the time derivative of the phase of the order parameter. Thus, as shown in Fig. 4(d), one expects the position of the single-particle gap to oscillate in momentum at the frequency of a coherently excited phase mode.

Although the explanation discussed so far is by far the most likely explanation, we note that the proposed parametric amplification mechanism might also apply to anharmonic coupling between amplitudons and other linearly dispersing excitations, such as for example acoustic phonons at $\omega_A/2$. In order to differentiate between the two scenarios, one should compare the strengths of the respective anharmonic coupling, as the strongest of the two dominate the parametric amplification process.

Considering a Free Energy surface that deviates from the idealized case depicted in Fig. 4(a), for example because of structural defects or near-commensurability effects [2], a coupling between amplitude and phase of the order parameter seems likely. Coupling to acoustic phonons would presumably involve an associated optical distortion of the lattice and the anharmonic decay into acoustic phonons.

Furthermore, as illustrated in Fig. 4(d), although coupling of phase mode to the electronic structure is well understood, acoustic phonons would couple to the band structure by weaker deformation potential [38], making it difficult to justify gap modulations of the size observed here.

Finally, we recall that we observe oscillations of the bonding band at the frequency of the amplitudon, and of the anti-bonding band at the frequency of the phason. The reason for such selective coupling is not uniquely understood, but we speculate that this may be related to the difference in orbital character between the two bands. As the anti-bonding band results from strongly hybridized Mo 4d and O 2p orbitals, whereas the bonding band is dominated by Mo 4d orbitals [39, 40], enhanced sensitivity to the motion of O and Mo atoms is expected in the two bands.

In summary, time-resolved ARPES with femtosecond EUV pulses is used to measure the ultrafast evolution of the electronic structure of blue bronze far from equilibrium. Impulsive laser excitation melts the CDW gaps, and the subsequent relaxation is accompanied by rigid momentum-dependent oscillations of the bands. As expected, the bonding band oscillates at $\omega_A$ = 1.7 THz, the frequency of the Raman-active amplitude mode. Most surprisingly, the anti-bonding band oscillates at $\omega_A/2$ = 0.8 THz. We assign this coherent oscillation to the Raman-inactive phase mode, which cannot be driven directly by displacing the atoms, but is presumably excited by parametric amplification of phase fluctuations, made possible by the anharmonic coupling between the amplitude and phase of the CDW order parameter.

In the future, similar time-resolved ARPES experiments should allow for a systematic study of similar parametric amplification processes in other complex materials, including, for example, coherent spin density wave excitations [41].

We thank D. Rice for technical support, and D.X. Mou for help with the crystal structure plots. Z.-A. Xu acknowledges support from the National Science Foundation of China. This research is supported by the EC Seventh Framework programme via Laserlab Europe.

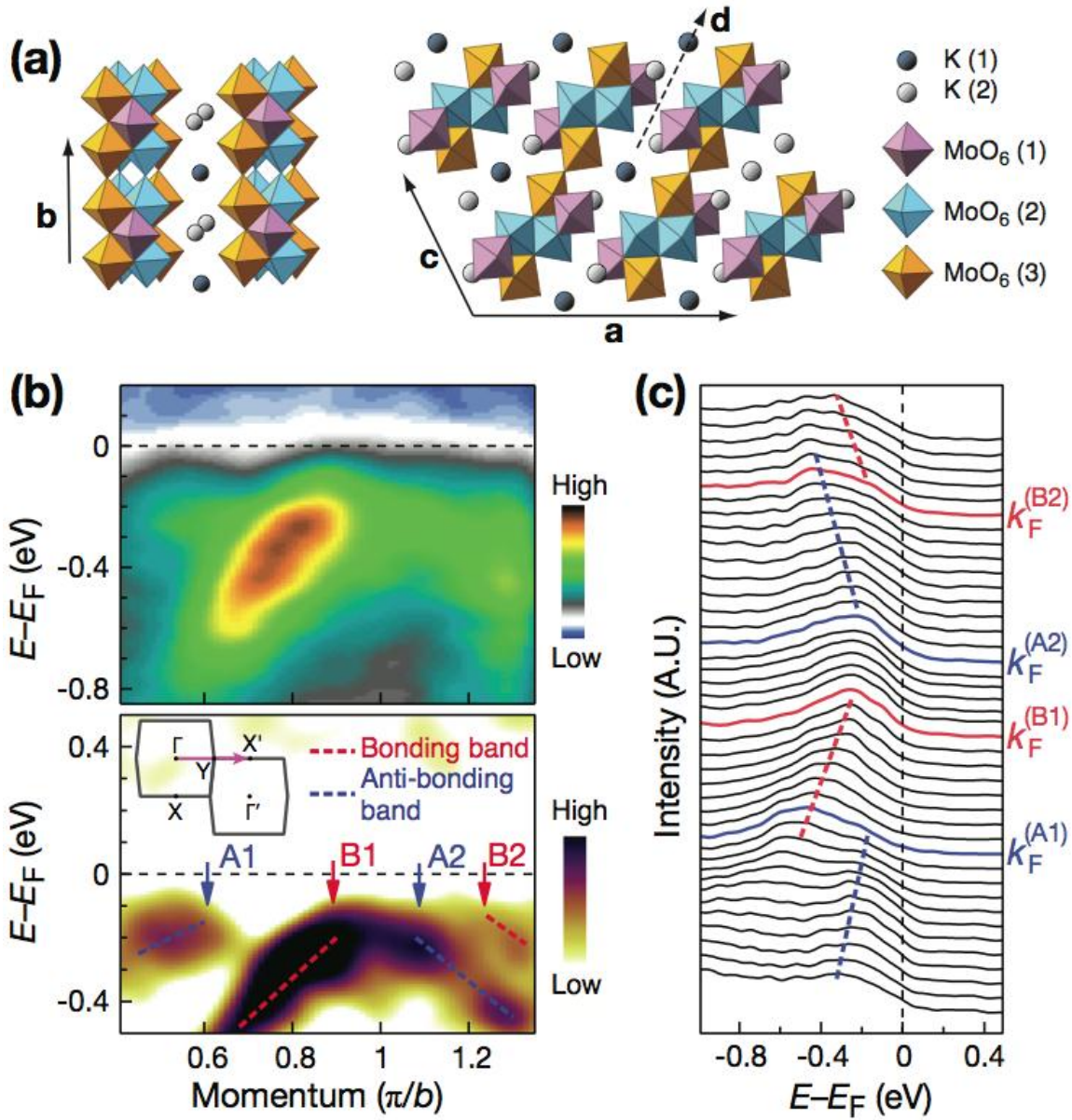

Fig. 1: Crystal structure and static band properties of $K_{0.3}MoO_3$. (a) Linear chains of $MoO_6$ octahedra form slabs, which make up the *b-d* cleavage plane. (b) Upper panel: static ARPES intensity map at *t* = -50 fs and 20 K, as a function of binding energy and momentum. To bring out the underlying band structure, we also plot $\partial^2 I/\partial E^2$ (lower panel). The dashed blue and red lines locate the anti-bonding (A1 and A2) and bonding (B1 and B2) bands. Vertical arrows mark the different Fermi wave-vectors. The photoelectron momentum is measured along the cut direction indicated on the zone map shown in the inset. (c) Individual EDCs used to produce the intensity map. EDCs at the Fermi momenta are highlighted and labeled.

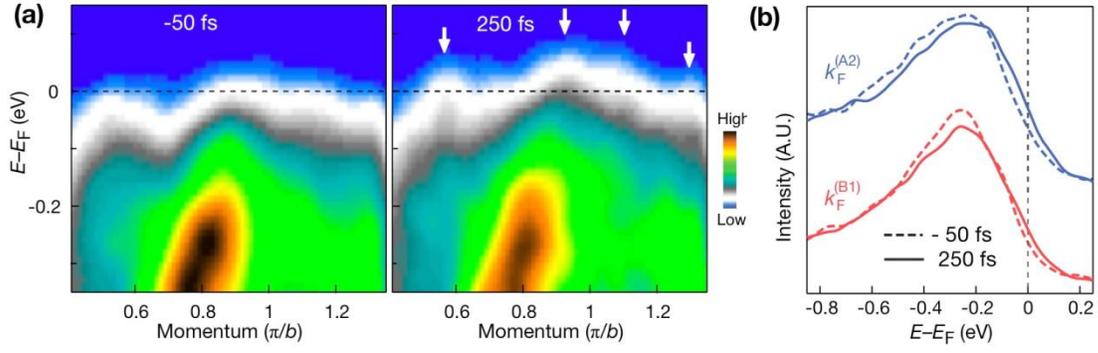

Fig. 2: Response to impulsive excitation by a femtosecond laser pulse. (a) Photoemission intensity maps at -50 fs and 250 fs, shown on the same color scale. The arrows mark the centres of increased intensity above $E_F$. (b) Photoinduced changes in the EDCs at $k_F^{(B1)}$ and $k_F^{(A2)}$, with intensity transferred from the valence bands to $E_F$ as the CDW phase melts in response to the excitation pulse.

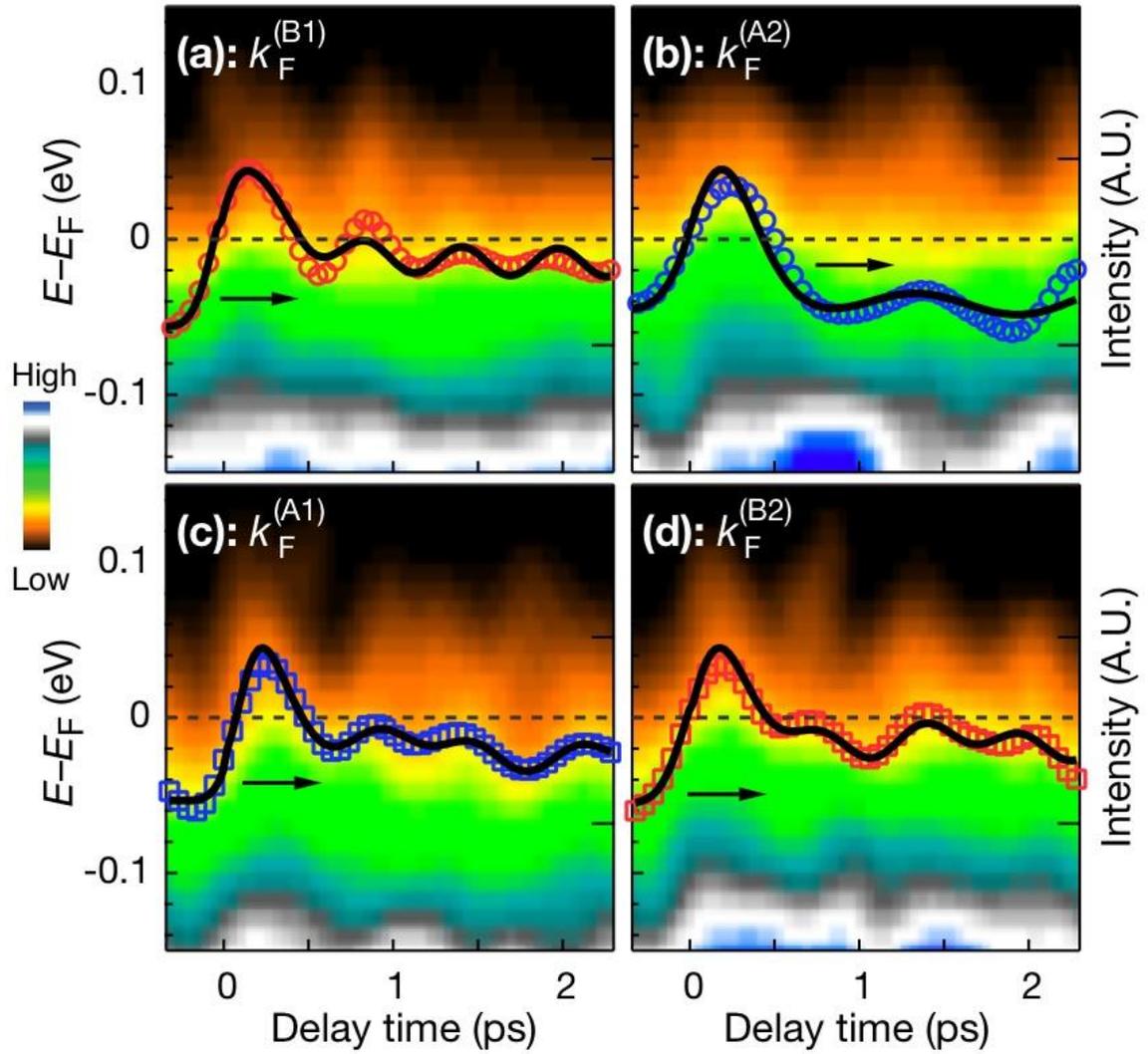

Fig. 3: Time evolution of electronic structure near the gap edge for each Fermi wavevector, as labeled. Symbols: intensity near $E_F$, integrated from -50 to +100 meV. Solid lines: fits to the integrated intensity. The model contains an exponential decay plus a sine function, with frequencies of 1.7 THz (a), 0.8 THz (b), or both (c, d).

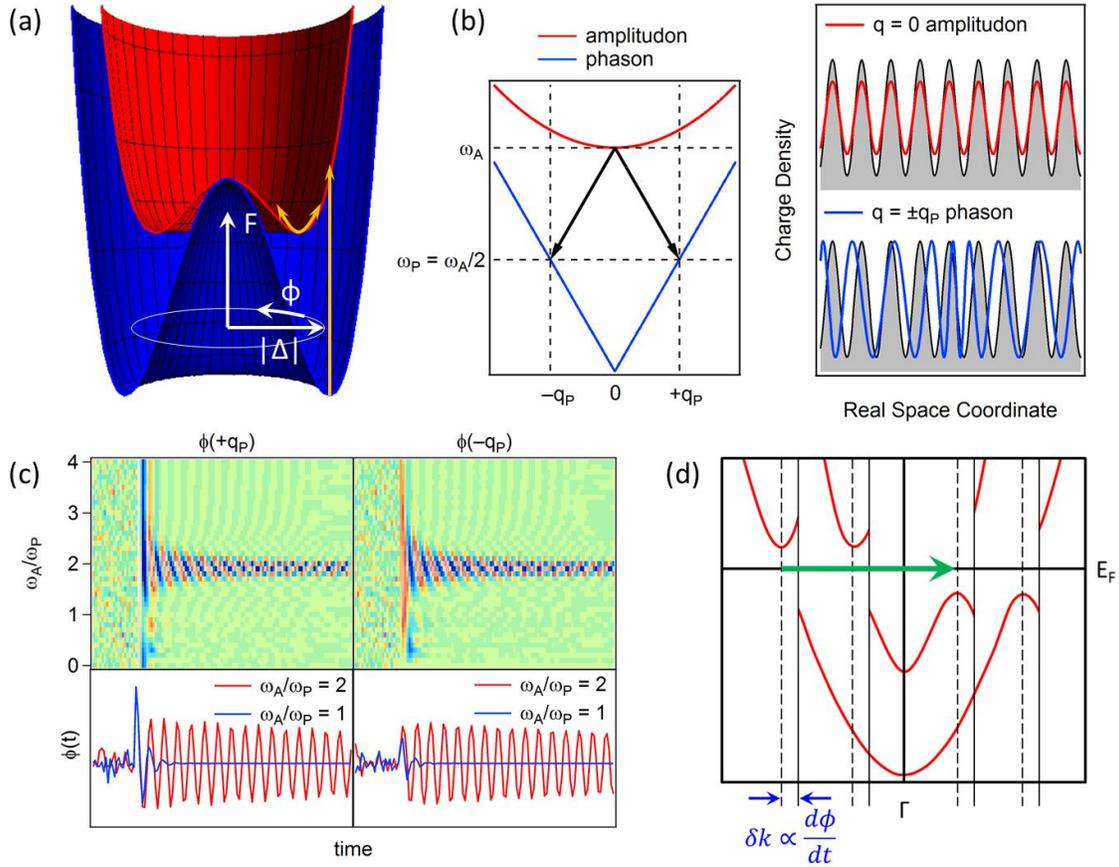

FIG. 4: Collective excitations of the CDW state. (a) Free energy of the ground state (blue) and the excited state (red) as a function of amplitude $|\Delta|$ and phase $\Phi$ of the complex order parameter. As the minimum of the free energy in the excited state is located at a different $|\Delta|$ compared to the ground state, amplitude oscillations are excited (yellow arrows). (b) Idealized dispersion and snapshots of the charge density of an amplitude mode (red) and phase mode (blue). Arrows indicate a parametric generation process from $\omega_A(0)$ to $\omega_P(\pm q_P) = \omega_A(0)/2$. (c) Coherent phase mode oscillations obtained by solving the coupled differential equations given in the text for different ratios of $\omega_A/\omega_P$. (d) Dispersion (red curves) and Fermi surface nesting (green arrow) in the presence of a time-dependent phase.